\documentclass[9pt,twocolumn,twoside]{pnas-new-edit}
\templatetype{pnasresearcharticle} % Choose template 
\setboolean{displaywatermark}{false}

\makeatother 

\RequirePackage{xcolor} % also loaded by tikz
\definecolor{darkgreen}{rgb}{0, 0.4, 0} 
\definecolor{midgreen}{rgb}{0.5, 0.8, 0.5}
\definecolor{darkred}{rgb}{0.5, 0, 0}  %% now a bit darker
\definecolor{darkblue}{rgb}{0, 0, 0.5}

\RequirePackage{hyperref}

\usepackage{microtype}

\def\bfseries{\fontseries \bfdefault \selectfont \boldmath}

\newcommand{\popt}{p_\star}
\newcommand{\pjeff}{p_\mathrm{J}}
\newcommand{\fkl}{f_\mathrm{KL}}
\newcommand{\MI}{\mathrm{MI}}

\newcommand{\atoms}{K} %% was N but people confuse this with m
\newcommand{\slope}{\zeta} %% was \alpha
\newcommand{\algtime}{\tau} %% was t but we use that elsewhere, maybe OK 

\newcommand{\figone}[1]{\hyperref[fig:one]{1#1}}
\newcommand{\figtwo}[1]{\hyperref[fig:two]{2#1}}

\newcommand{\figthree}[1]{\hyperref[fig:three]{4#1}} %% sorry!
\newcommand{\figfour}[1]{\hyperref[fig:four]{5#1}}

\newcommand{\figsix}[1]{\hyperref[fig:six]{S1#1}}

\allowdisplaybreaks

\makeatletter

\title{Maximizing the information learned from finite data selects a simple model}

\author[a]{Henry H. Mattingly}
\author[b]{Mark K. Transtrum} 
\author[c,1]{Michael C. Abbott}
\author[d,1]{Benjamin B. Machta}

\affil[a]{Lewis-Sigler Institute and Department of Chemical and Biological
Engineering, Princeton University, Princeton, NJ 08544, USA}
\affil[b]{Department of Physics and Astronomy, Brigham Young University, Provo,
Utah 84602, USA}
\affil[c]{Marian Smoluchowski Institute of Physics, Jagiellonian University, Ulica {\L}ojasiewicza
11, 30-348 Krak\'{o}w, Poland}
\affil[d]{Lewis-Sigler Institute and Department of Physics, Princeton University,
Princeton, NJ 08544, USA}

\leadauthor{Mattingly}
\significancestatement{

Most physical theories are effective theories, descriptions at the scale visible to our experiments which ignore microscopic details. Seeking general ways to motivate such theories, we find an information theory perspective: if we select the model which can learn as much information as possible from the data, then we are naturally led to a simpler model, by a path independent of concerns about overfitting. This is encoded as a Bayesian prior which is nonzero only on a subspace of the original parameter space. We differ from earlier prior selection work by not considering infinitely much data. Having finite data is always a limit on the resolution of an experiment, and in our framework this selects how complicated a theory is appropriate. 

}

\authorcontributions{H.H.M. and M.C.A. performed the numerical experiments. H.H.M., M.K.T., M.C.A. and B.B.M. designed the research, interpreted results, and contributed to writing. M.C.A. and B.B.M. led the writing of the paper.}
\correspondingauthor{\textsuperscript{1}To whom correspondence should be addressed. E-mail: abbott@th.if.uj.edu.pl and\\ benjamin.machta@yale.edu}

\authordeclaration{Present affiliation  for H.H.M.: 
Department of Molecular Cellular and Developmental Biology, Yale University, New Haven, CT, 06520 \medskip \\ 
Present affiliations for B.B.M: 
Department of Physics, Yale University, New Haven, CT, 06520 \\ Systems Biology Institute, Yale University, West Haven, CT, 06516
}

\keywords{Effective Theory $|$ Model Selection $|$ Renormalization Group $|$ Bayesian Prior Choice $|$ Information Theory  } 

\begin{abstract}
We use the language of uninformative Bayesian prior choice
to study the selection of appropriately simple effective models. We
advocate for the prior which maximizes the mutual information between
parameters and predictions, learning as much as possible from limited
data. When many parameters are poorly constrained by the available
data, we find that this prior puts weight only on boundaries of the
parameter manifold. Thus it selects a lower-dimensional effective
theory in a principled way, ignoring irrelevant parameter directions.
In the limit where there is sufficient data to tightly constrain any
number of parameters, this reduces to Jeffreys prior. But we argue
that this limit is pathological when applied to the hyper-ribbon parameter
manifolds generic in science, because it leads to dramatic dependence
on effects invisible to experiment. \end{abstract}

\newcommand{\mytoday}{7 February, 2018}

\dates{\href{http://www.pnas.org/content/early/2018/02/05/1715306115}{PNAS} (7 February 2018)
 + \href{http://www.pnas.org/content/pnas/suppl/2018/02/05/1715306115.DCSupplemental/pnas.201715306SI.pdf}{SI} 
 $\approx$ \href{http://arxiv.org/abs/1705.01166}{arXiv:1705.01166}v3 \\
Earlier versions had the title ``Rational Ignorance: Simpler Models Learn More Information from Finite Data''}

\begin{document}

% Optional adjustment to line up main text (after abstract) of first page with line numbers, when using both lineno and twocolumn options.
% You should only change this length when you've finalised the article contents.
\verticaladjustment{-2pt}

\maketitle
\thispagestyle{empty}

\ifthenelse{\boolean{shortarticle}}{\ifthenelse{\boolean{singlecolumn}}{\abscontentformatted}{\abscontent}}{}

% If your first paragraph (i.e. with the \dropcap) contains a list environment (quote, quotation, theorem, definition, enumerate, itemize...), the line after the list may have some extra indentation. If this is the case, add \parshape=0 to the end of the list environment.
\dropcap{P}hysicists prefer simple models not because nature is simple,
but because most of its complication is usually irrelevant. Our most
rigorous understanding of this idea comes from the Wilsonian renormalization
group \cite{Kadanoff:1966wm,Wilson:1971bg,Cardy:1996th}, which describes
mathematically the process of zooming out and losing sight of microscopic
details. These details only influence the effective theory which describes
macroscopic observables through a few relevant parameter combinations,
such as the critical temperature, or the proton mass. The remaining
irrelevant parameters can be ignored, as they are neither constrained
by past data nor useful for predictions. Such models can now be understood
as part of a large class called sloppy models \cite{Waterfall:2006fc,Gutenkunst:2007gl,Transtrum:2010ci,Transtrum:2011de,Machta:2013ga,Transtrum:2015hm,OLeary:2015ip,Niksic:2016qau,Raman:2016oes,Bohner:2017es,Raju:2017ty},
whose usefulness relies on a similar compression of a large microscopic
parameter space down to just a few relevant directions.

This justification for model simplicity is different from the one
more often discussed in statistics, motivated by the desire to avoid
overfitting \cite{Akaike:1974ih,Sugiura:1978be,Balasubramanian:1997di,Myung:2000in,Spiegelhalter:2002ii,Watanabe:2010uh,Wiggins:2016tb}.
Since irrelevant parameters have an almost invisible effect on predicted
data, they cannot be excluded on these grounds. Here we motivate their
exclusion differently: we show that simplifying a model can often
allow it to extract more information from a limited data set, and
that this offers a guide for choosing appropriate effective theories.

We phrase the question of model selection as part of the choice of
a Bayesian prior on some high-dimensional parameter space. In a set
of nested models, we can always move to a simpler model by using a
prior which is nonzero only on some subspace. Recent work has suggested
that interpretable effective models are typically obtained by taking
some parameters to their limiting values, often $0$ or $\infty$,
thus restricting to lower-dimensional boundaries of the parameter
manifold \cite{Transtrum:2014hr}.

Our setup is that we wish to learn about a theory by performing some
experiment which produces data $x\in X$. The theory and the experiment
are together described by a probability distribution $p(x\vert\theta)$,
for each value of the theory's parameters $\theta\in\Theta$. This
function encodes both the quality and quantity of data to be collected.

\begin{figure}
\centering \includegraphics[width=0.99\columnwidth]{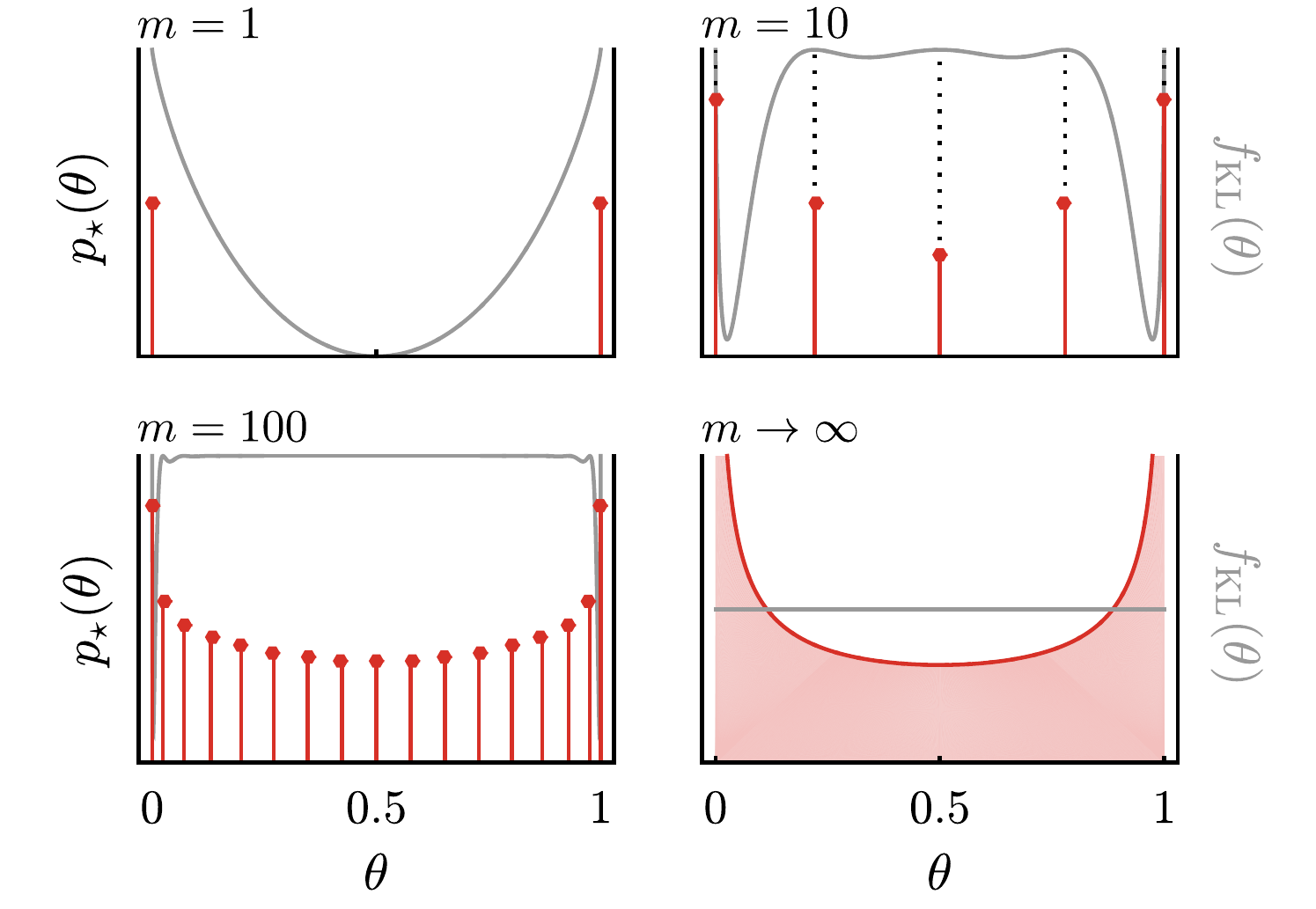}

\caption{\textbf{Optimal priors for the Bernoulli model (\ref{eq:p-cond-coins}).}
Red lines indicate the positions of delta functions in $p_{\star}(\theta)$,
which are at the maxima of $f_{\mathrm{KL}}(\theta)$, \eqref{eq:defn-MI-with-f}.
As $m\to\infty$ these coalesce into Jeffreys prior $p_{\mathrm{J}}(\theta)$.
\label{fig:one}}
\end{figure}

\begin{figure}
\centering \includegraphics[width=0.90\columnwidth]{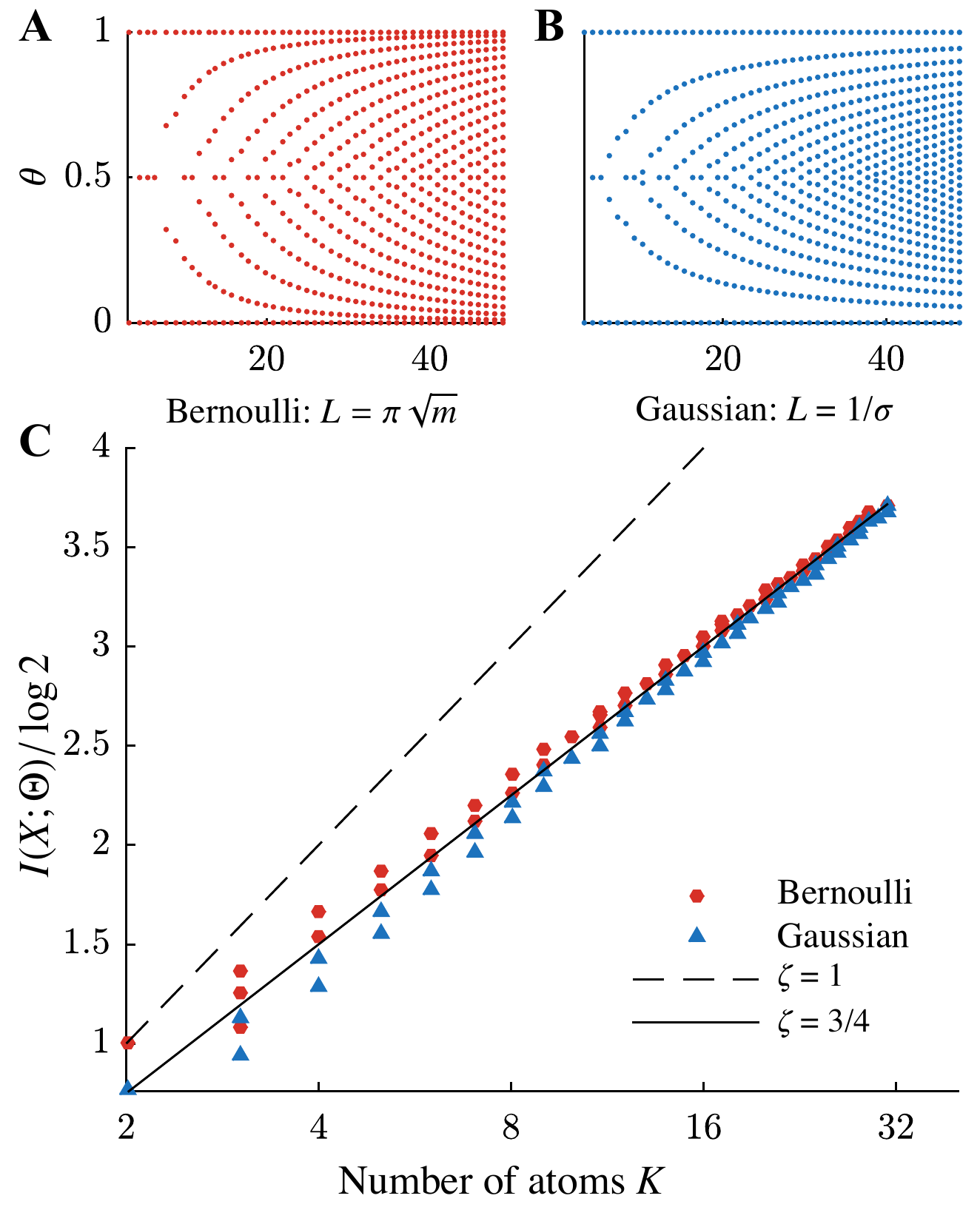}

\caption{\textbf{Behavior of $p_{\star}(\theta)$ with increasing Fisher length.}
Panels A and B show the atoms of $p_{\star}(\theta)$ for the two
one-dimensional models as $L$ is increased (i.e. we perform more
repetitions $m$ or have smaller noise $\sigma$). Panel C shows the
scaling of the mutual information (in bits) with the number of atoms
$K$.The dashed line is the bound $\mathrm{MI}\leq\log K$, and the
solid line is the scaling law $\mathrm{MI}\sim3/4\log K$% 
% leading to \eqref{eq:scaling-law-rho}%
. \label{fig:two}}
\end{figure}

The mutual information between the parameters and their expected data
is defined as $\MI=I(X;\Theta)=S(\Theta)-S(\Theta\vert X)$, where
$S$ is the Shannon entropy \cite{Shannon:1948wk}. The MI thus quantifies
the information which can be learned about the parameters by measuring
the data, or equivalently, the information about the data which can
be encoded in the parameters \cite{Lindley:1956bj,Renyi:1967tt}.
Defining $\popt(\theta)$ by maximizing this, we see:
\begin{enumerate}
\item The prior $\popt(\theta)$ is almost always discrete \cite{Farber:1967us,Smith:1971kt,Fix:1978vk,Berger:1988vs,Zhang:1994ui},
with weight only on a finite number $\atoms$ of points, or atoms
(Figures \ref{fig:one} and \ref{fig:BA-new}): $\popt(\theta)=\sum_{a=1}^{\atoms}\lambda_{a}\delta(\theta-\theta_{a}).$ 
\item When data is abundant, $\popt(\theta)$ approaches Jeffreys prior
$\pjeff(\theta)$ \cite{Bernardo:1979uq,Clarke:1994gw,Scholl:1998kq}.
As this continuum limit is approached, the proper spacing of the atoms
shrinks as a power law (Figure \ref{fig:two}). \label{enu:As-data-becomes-plentiful} 
\item When data is scarce, most atoms lie on boundaries of parameter space,
corresponding to effective models with fewer parameters (Figure \ref{fig:three}).
The resulting distribution of weight along relevant directions is
much more even than that given by Jeffreys prior (Figure \ref{fig:four}).
\label{enu:When-data-is-sparse} 
\end{enumerate}
After some preliminaries, we demonstrate these properties in three
simple examples, each a stylized version of a realistic experiment.
To see the origin of discreteness, we study the bias of an unfair
coin and the value of a single variable corrupted with Gaussian noise.
To see how models of lower dimension arise, we then study the problem
of inferring decay rates in a sum of exponentials.

In the Appendix we discuss the alorithms used for finding $\popt(\theta)$
(Figures \ref{fig:BA-new} and \ref{fig:BA-convergence-2D}), and
we apply some more traditional model selection tools to the sum of
exponentials example (Figure \ref{fig:six}).

\section*{Priors and Geometry}

Bayes' theorem tells us how to update our knowledge of $\theta$
upon observing data $x$, from prior $p(\theta)$ to posterior $p(\theta|x)=p(x|\theta)\:p(\theta)/p(x)$,
where $p(x)=\int d\theta\ p(\theta)\ p(x|\theta)$. In the absence
of better knowledge we must pick an uninformative prior which codifies
our ignorance. The naive choice of a flat prior $p(\theta)=\text{const}.$
has undesirable features, in particular making $p(x)$ depend on the
choice of parameterization, through the measure $d\theta$.

Jeffreys prior $\pjeff(\theta)$ is invariant under changes of parameterizaion,
because it is constructed from some properties of the experiment 
\cite{Jeffreys:1946jf}. This $\pjeff(\theta)\propto\sqrt{\det g_{\mu\nu}}$
is, up to normalization, the volume form arising from the Fisher information
metric (FIM, often -matrix): 
\[
g_{\mu\nu}(\vec{\theta})=\int dx\:p(x|\vec{\theta})\:\frac{\partial\log p(x|\vec{\theta})}{\partial\theta^{\mu}}\:\frac{\partial\log p(x|\vec{\theta})}{\partial\theta^{\nu}}.
\]
This Riemannian metric defines a reparameterization-invariant distance
between points, $ds^{2}=\sum_{\mu,\nu=1}^{D}g_{\mu\nu}\,d\theta^{\mu}d\theta^{\nu}$.
It measures the distinguishability of the data which $\theta$ and
$\theta+d\theta$ are expected to produce, in units of standard deviations.
Repeating an (identical and independently distributed) experiment
$m$ times means considering $p^{m}(\vec{x}|\theta)=\prod_{j=1}^{m}p(x_{j}|\theta)$,
which leads to metric $g_{\mu\nu}^{m}(\theta)=m\:g_{\mu\nu}(\theta)$.
However the factor $m^{D/2}$ in the volume is lost by normalizing
$\pjeff(\theta)$. Thus Jeffreys prior depends on the type of experiment,
but not the quantity of data.

Bernardo defined a prior $\popt(\theta)$ by maximizing the mutual
information between parameters $\Theta$ and the expected data $X^{m}$
from $m$ repetitions, and then a reference prior by taking the limit
$m\to\infty$ \cite{Bernardo:1979uq,Berger:1988vs}. Under certain
benign assumptions, this reference prior is exactly Jeffreys prior
\cite{Bernardo:1979uq,Clarke:1994gw,Scholl:1998kq}, providing an
alternative justification for $\pjeff(\theta)$.

We differ in taking seriously that the amount of data collected is
always finite.\footnote{Interned for five years, John Kerrich only flipped his coin $10^{4}$
times \cite{Kerrich:1946wy}. With computers we can do better, but
even the LHC only generated about $10^{18}$ bits of data \cite{cern:petabytes}.} Besides being physically unrealistic, the limit $m\to\infty$ is
pathological both for model selection and prior choice. In this limit
any number of parameters can be perfectly inferred, justifying an
arbitrarily complicated model. In addition, in this limit the posterior
$p(\theta|x)$ becomes independent of any smooth prior.\footnote{For simplicity we consider only regular models, i.e. we assume all
parameters are structurally identifiable.}

Geometrically, the defining feature of sloppy models is that they
have a parameter manifold with hyper-ribbon structure \cite{Machta:2013ga,Transtrum:2010ci,Transtrum:2011de,Transtrum:2015hm}:
there are some long directions (corresponding to $d$ relevant, or
stiff, parameters) and many shorter directions ($D-d$ irrelevant,
or sloppy, parameter combinations). These lengths are often estimated
using the eigenvalues of $g_{\mu\nu}$, and have logarithms that are
roughly evenly-spaced over many orders of magnitude \cite{Waterfall:2006fc,Gutenkunst:2007gl}.
The effect of coarse-graining is to shrink irrelevant directions (here
using the technical meaning of irrelevant: a parameter which shrinks
under renormalization group flow) while leaving relevant directions
extended, producing a sloppy manifold \cite{Machta:2013ga,Raju:2017ty}.
By contrast the limit $m\to\infty$ has the effect of expanding all
directions, thus erasing the distinction between directions longer
and shorter than the critical length scale of (approximately) one
standard deviation.

On such a hyper-ribbon, Jeffreys prior has an undesirable feature:
since it is constructed from the $D$-dimensional notion of volume,
its weight along the relevant directions always depends on the volume
of the $D-d$ irrelevant directions. This gives it extreme dependence
on which irrelevant parameters are included in the model.\footnote{See Figure \ref{fig:four} for a demonstration of this point. For
another example, consider a parameter manifold $\Theta$ which is
a cone, with Fisher metric $ds^{2}=(50\,d\vartheta)^{2}+\vartheta^{2}d\Omega_{n}^{2}/4$:
there is one relevant direction $\vartheta\in[0,1]$ of length $L=50$,
and $n$ irrelevant directions forming a sphere of diameter $\vartheta$.
Then the prior on $\vartheta$ alone implied by $\pjeff(\vec{\theta})$
is $p(\vartheta)=(n+1)\vartheta^{n}$, putting most of the weight
near to $\vartheta=1$, dramatically so if $n=D-d$ is large. But
since only the relevant direction is visible to our experiment, the
region $\vartheta\approx0$ ought to be treated similarly to $\vartheta\approx1$.
The prior $\popt(\vec{\theta})$ has this property. } The optimal prior $\popt(\theta)$ avoids this dependence because
it is almost always discrete, at finite $m$.\footnote{We offer both numerical and analytic arguments for discreteness below.
The exception to discreteness is that if there is an exact continuous
symmetry, $\popt(\theta)$ will be constant along it. For example
if our Gaussian model \eqref{eq:p-cond-gaussian} is placed on a circle
(identifying both $\theta\sim\theta+1$ and $x\sim x+1)$ then the
optimum prior is a constant.} It puts weight on a set of nearly distinguishable points, closely
spaced along the relevant directions, but ignoring the irrelevant
ones. Yet being the solution to a reparameterization-invariant optimization
problem, the prior $\popt(\theta)$ retains this good feature of $\pjeff(\theta)$.

Maximizing the mutual information was originally done to calculate
the capacity of a communication channel, and we can borrow techniques
from rate-distortion theory here: the algorithms we use were developed
there \cite{Arimoto:1972jz,Blahut:1972ed}, and the discreteness we
exploit was discovered several times in engineering  \cite{Farber:1967us,Smith:1971kt,Fix:1978vk,Rose:1994km}.
In statistics, this problem is more often discussed as an equivalent
minimax problem \cite{Haussler:1997fa}. Discreteness was also observed
in other minimax problems \cite{Ghosh:1964ga,Casella:1981ex,Feldman:1991ba},
and later in directly maximising mutual information \cite{Berger:1988vs,Zhang:1994ui,Scholl:1998kq,Chen:2010us}.
However it does not seem to have been seen as useful, and none of
these papers explicitly find discrete priors in dimension $D>1$,
which is where we see attractive properties. 
Discreteness has been useful, although for different reasons, in the idea of rational inattention in economics \cite{Sims:2006gu,Jung:2015uy}. 
There, market actors have a finite bandwidth for news, and this drives
them to make discrete choices despite all the dynamics being continuous.
 Rate-distortion theory has also been useful in several areas of biology
\cite{Laughlin:1981hx,Tkacik:2008dq,Petkova:2016up}, and discreteness
emerges in a recent theoretical model of the immune system \cite{Mayer:2015ce}.

We view this procedure of constructing the optimal prior as a form
of model selection, picking out the subspace of $\Theta$ on which
$\popt(\theta)$ has support. This depends on the likelihood function
$p(x\vert\theta)$ and the data space $X$, but not on the observed
data $x$. In this regard it is closer to Jeffreys' perspective on
prior selection  than to tools like the information criteria and Bayes
factors,  which are employed at the stage of fitting to data. We discuss
this difference at more length in the Appendix. 

\section*{One-Parameter Examples}

We begin with some problems with a single bounded parameter,
of length $L$ in the Fisher metric. These tractable cases illustrate
the generic behaviour along either short (irrelevant) or long (relevant,
$L\gg1$) parameter directions in higher-dimensional examples.

Our first example is the Bernoulli problem, in
which we wish to determine the probability $\theta\in[0,1]$ that
an unfair coin gives heads, using the data from $m$ trials. It is
sufficient to record the total number of heads $x$, which occurs
with probability 
\begin{equation}
p(x|\theta)=\frac{m!}{x!(m-x)!}\,\theta^{x}(1-\theta)^{m-x}.\label{eq:p-cond-coins}
\end{equation}
This gives $g_{\theta\theta}=\frac{m}{\theta(1-\theta)}$, thus $\pjeff(\theta)=[\pi\sqrt{\theta(1-\theta)}]^{-1}$,
and proper parameter space length $L=\int\sqrt{ds^{2}}=\pi\sqrt{m}$.

In the extreme case $m=1$, the optimal prior is two delta functions,
$\popt(\theta)=\tfrac{1}{2}\delta(\theta)+\tfrac{1}{2}\delta(\theta-1)$,
and $\MI=\log2$, exactly one bit \cite{Berger:1988vs,Zhang:1994ui,Scholl:1998kq}.
Before an experiment that will only ever be run once, this places
equal weight on both outcomes; afterwards it records the outcome.
As $m$ increases, weight is moved from the boundary onto interior
points, which increase in number, and ultimately approach the smooth
$\pjeff(\theta)$: see Figures \ref{fig:one} and \figtwo{A}.

\begin{figure}
\centering \includegraphics[width=0.95\columnwidth]{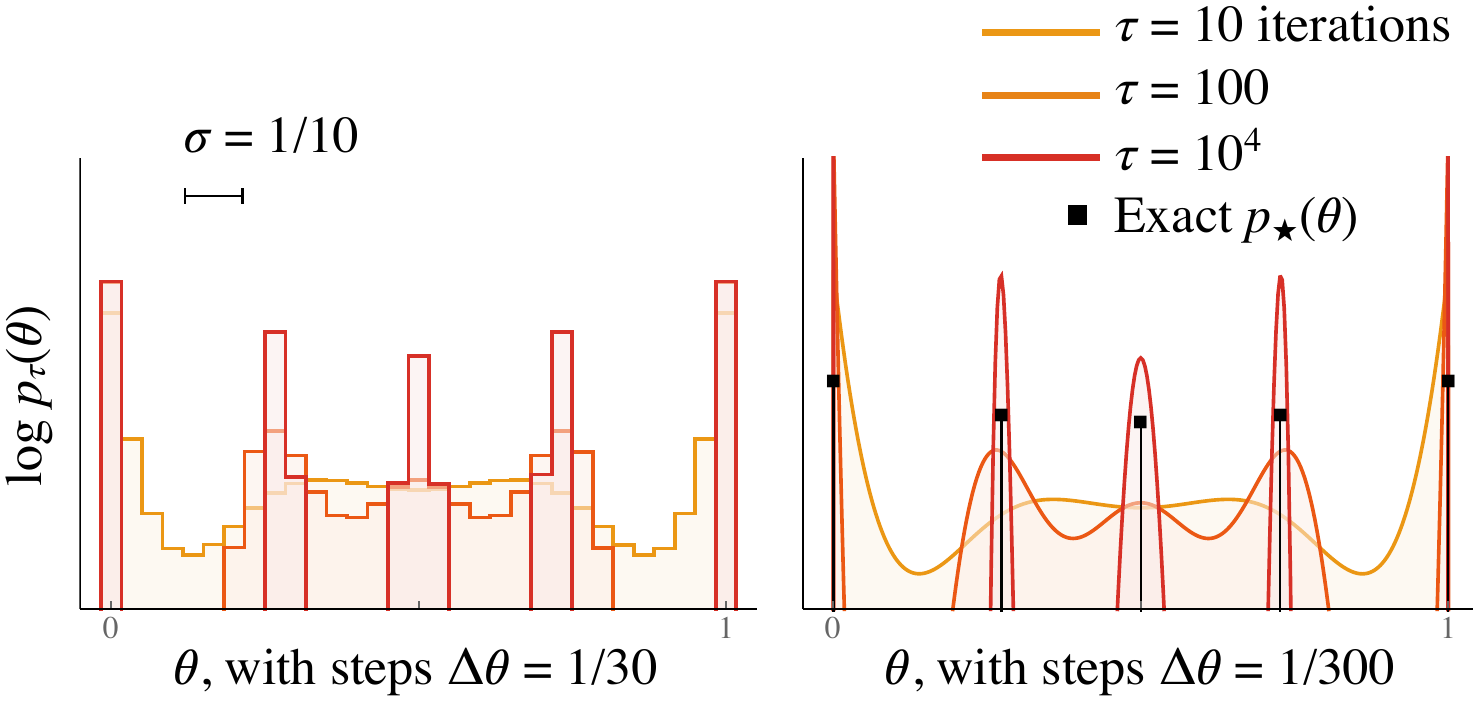} \caption{\textbf{Convergence of the  Blahut\textendash Arimoto algorithm.}
This is for the one-parameter Gaussian model \eqref{eq:p-cond-gaussian}
with $L=10$ (comparable to $m=10$ in Figure \ref{fig:one}). On
the right $\theta$ is discretized into ten times as many points,
but $p_{\tau}(\theta)$ clearly converges to the same five delta functions.
\label{fig:BA-new}}
\end{figure}

Similar behavior is seen in a second example, in which we measure
one real number $x$, normally distributed with known $\sigma$ about
the parameter $\theta\in[0,1]$: 
\begin{equation}
p(x\vert\theta)=\frac{1}{\sqrt{2\pi}\sigma}e^{-(x-\theta)^{2}/2\sigma^{2}}.\label{eq:p-cond-gaussian}
\end{equation}
Repeated measurements are equivalent to smaller $\sigma$ (by $\sigma\to\sigma/\sqrt{m}$),
so we fix $m=1$ here. The Fisher metric is $g_{\theta\theta}=1/\sigma^{2}$,
thus $L=1/\sigma$. An optimal prior is shown in Figure \ref{fig:BA-new},
and in Figure \figfour{A} along with its implied distribution of
expected data. This $p(x)$ is similar to that implied by Jeffreys
prior, here $\pjeff(\theta)=1$.

We calculated $\popt(\theta)$ numerically in two ways. After discretizing
both $\theta$ and $x$, we can use the Blahut\textendash Arimoto
(BA) algorithm \cite{Arimoto:1972jz,Blahut:1972ed}. This converges
to the global maximum, which is  a discrete distribution: see Figure
\ref{fig:BA-new}. Alternatively, using our knowledge that $\popt(\theta)$
is discrete, we can instead adjust the positions $\theta_{a}$ and
weights $\lambda_{a}$ of a finite number of atoms. See the %methods section 
Supplement for more details.

To see analytically why discreteness arises, we write the mutual information
as 
\begin{align}
 & \MI=I(X;\Theta)= \int d\theta\:p(\theta)\:\fkl(\theta),\label{eq:defn-MI-with-f}\\
\fkl(\theta) & =D_{\mathrm{KL}}\left[p(x\vert\theta)\middle\|p(x)\vphantom{1^{1}}\right]=\int dx\:p(x\vert\theta)\:\log\frac{p(x\vert\theta)}{p(x)}\nonumber 
\end{align}
where $D_{\mathrm{KL}}$ is the Kullback\textendash Leibler divergence.\footnote{The function $\fkl(\theta)$ is sometimes called the Bayes risk, as
it quantifies how poorly the prior will perform if $\theta$ turns
out to be correct. One of the problems equivalent to maximising the
mutual information \cite{Haussler:1997fa} is the minimax problem
for this (see also Figure \ref{fig:one}):
%\begin{align*}
\[
\max_{p(\theta)}I(X;\Theta)  =\min_{p(\theta)}\max_{\theta}\fkl(\theta)
  =\min_{q(x)}\max_{p(\theta)}\textstyle\int d\theta\:p(\theta)\:D_{\mathrm{KL}}\left[p(x\vert\theta)\middle\|q(x)\right].
\]
%\end{align*}
The distributions we call expected data $p(x)$ are also known as
Bayes strategies, i.e. distributions on $X$ which are the convolution
of the likelihood $p(x\vert\theta)$ with some prior $p(\theta)$.
The optimal $q(x)$ from this third formulation (with $\min_{q(x)}\ldots$)
can be shown to be such a distribution \cite{Haussler:1997fa}.} Maximizing $\MI$ over all functions $p(\theta)$ with $\int d\theta\:p(\theta)=1$
gives $\fkl(\theta)=$~const. But the maximizing function will not,
in general, obey $p(\theta)\geq0$. Subject to this inequality $\popt(\theta)$
must satisfy 
\[
\left\{ \popt(\theta)>0\text{, }\fkl(\theta)=\MI\right\} \;\text{or}\;\left\{ \popt(\theta)=0\text{, }\fkl(\theta)<\MI\right\} 
\]
at every $\theta$. With finite data $\fkl(\theta)-\MI$ must be an
analytic function of $\theta$, and therefore must be smooth with
a finite numbers of zeros, corresponding to the atoms of $\popt(\theta)$
(Figure \figone{A}). See \cite{Fix:1978vk,Berger:1988vs,Jung:2015uy}
for related arguments for discreteness, and \cite{Ghosh:1964ga,Casella:1981ex,Feldman:1991ba}
for other approaches.

The number of atoms occurring in $\popt(\theta)$ increases as the
data improves. For $\atoms$ atoms there is an absolute bound $\MI\leq\log\atoms$,
saturated if they are perfectly distinguishable. In Figure \figtwo{C}
we observe that the optimal priors instead approach a line $\MI\to\slope\log\atoms$,
with slope $\slope\approx0.75$. At large $L$ the length of parameter
space is proportional to the number of distinguishable points, hence
$\MI\to\log L$. Together these imply $\atoms\sim L^{1/\slope}$,
and so the average number density of atoms grows as 
\begin{equation}
\rho_{0}=\atoms/L\sim L^{1/\slope-1}\approx L^{1/3},\qquad L\gg1.\label{eq:scaling-law-rho}
\end{equation}
Thus the proper spacing between atoms shrinks to zero in the limit
of infinite data, i.e. neighboring atoms cease to be distinguishable.

To derive this scaling law analytically, in a related paper \cite{Abbott:2017oqw}
we consider a field theory for the number density of atoms, in which
the entropy density (omitting numerical factors) is 
\[
\mathcal{S}=\text{const.}-\smash{e^{-\rho^{2}}}[\rho^{4}(\rho')^{2}+1].
\]
From this we find $\slope=3/4$, which is consistent with both examples
presented above.

\section*{Multi-parameter Example}

 In the examples above, $\popt(\theta)$ concentrates weight
on the edges of its allowed domain when data is scarce (i.e. when
$m$ is small or $\sigma$ is large, hence $L$ is small). We next
turn to a multi-parameter model in which some parameter combinations
are ill-constrained, and where edges correspond to reduced models.

The physical picture is that we wish to determine the composition
of an unknown radioactive source, from data of $x_{t}$ Geiger counter
clicks at some times $t$. As parameters we have the quantities $A_{\mu}$
and decay constants $k_{\mu}$ of isotopes $\mu$. The probability
of observing $x_{t}$ should be a Poisson distribution (of mean $y_{t}$)
at each time, but we approximate these by Gaussians 
%of fixed $\sigma$ 
to write:\footnote{Using a normal distribution of fixed $\sigma$ here is what allows
the metric in \eqref{eq:metric-k-y} to be so simple. However the
qualitative behavior from the Poisson distribution is very similar.} 
\begin{equation}
p(\vec{x}\vert\vec{y})\propto \prod_{t}e^{-(x_{t}-y_{t})^{2}/2\sigma^{2}},\qquad y_{t}=\sum_{\mu}A_{\mu}e^{-k_{\mu}t}.\label{eq:forward-exp-model}
\end{equation}
Fixing $\sigma_{t}=\sigma=\text{const.}$ then brings us to a nonlinear
least-squares model of the type studied by \cite{Transtrum:2010ci,Transtrum:2011de}. 
This same model also arises in other contexts, such as the asymptotic
approach to equilibrium of many dynamical systems \cite{Strogatz:2007book}.

We can see the essential behavior with just two isotopes in fixed
quantities: $A_{\mu}=\tfrac{1}{2}$, thus $y_{t}=\tfrac{1}{2}(e^{-k_{1}t}+e^{-k_{2}t})$.
Measuring at only two times $t_{1}$ and $t_{2}$, we almost have
a two-dimensional version of \eqref{eq:p-cond-gaussian}, in which
the center of the distribution $\smash{\vec{y}=(y_{1},y_{2})}$ plays
the role of $\theta$ above. The mapping between $\smash{\vec{k}}$
and $\smash{\vec{y}}$ is shown in Figure \figthree{A}, fixing $t_{2}/t_{1}=e$.
The Fisher information metric is proportional to the ordinary Euclidean
metric for $\vec{y}$, but not for $\smash{\vec{k}}$: 
\begin{equation}
g_{\mu\nu}(\vec{k})=\frac{1}{\sigma^{2}}\sum_{t}\frac{\partial y_{t}}{\partial k_{\mu}}\frac{\partial y_{t}}{\partial k_{\nu}}\quad\negmedspace\Longleftrightarrow\negmedspace\quad g_{st}(\vec{y})=\frac{1}{\sigma^{2}}\delta_{st}.\label{eq:metric-k-y}
\end{equation}
Thus Jeffreys prior is a constant on the allowed region of the $\vec{y}$
plane.

\begin{figure}
\centering\includegraphics[width=0.99\columnwidth]{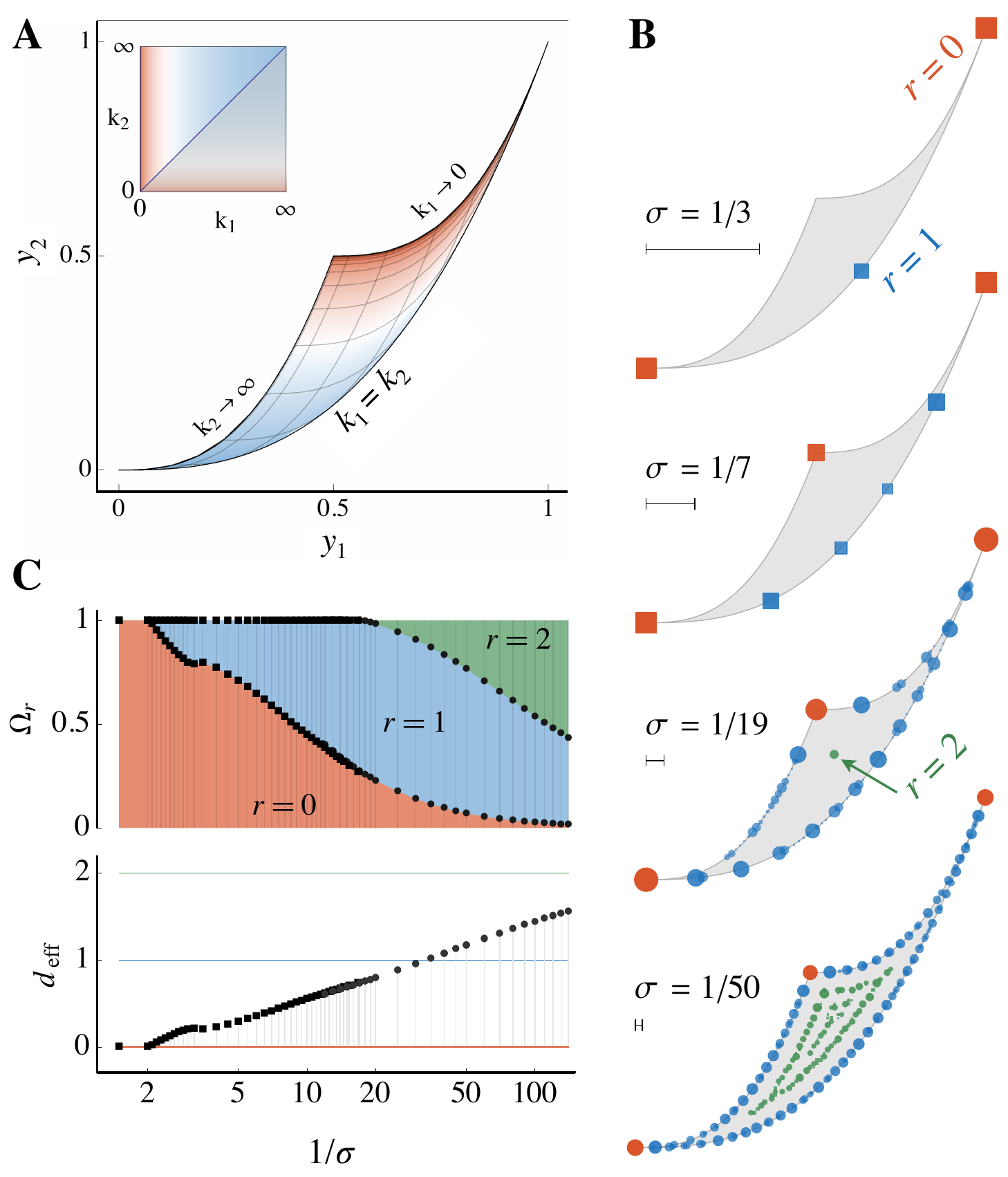}

\caption{\textbf{Parameters and priors for the exponential model (\ref{eq:forward-exp-model}).}
Panel A shows the area of the $\vec{y}$ plane covered by all decay
constants $k_{1},k_{2}\geq0$.  Panel B shows the positions of the
delta functions of the optimal prior $p_{\star}(\vec{y})$ for several
values of $\sigma$, with colors indicating the dimensionality $r$
at each point. Panel C shows the proportion of weight on these dimensionalities.
\label{fig:three} }
\end{figure}

Then we proceed to find the optimum $\popt(\vec{y})$ for this model,
shown in Figure \figthree{B} for various values of $\sigma$. When
$\sigma$ is large, this has delta functions only in two of the corners,
allowing only $k_{1},k_{2}=0$ and $k_{1},k_{2}=\infty$. As $\sigma$
is decreased, new atoms appear first along the lower boundary (corresponding
to the one-dimensional model where $k_{1}=k_{2}$) and then along
the other boundaries. At sufficiently small $\sigma$, atoms start
filling in the (two-dimensional) interior.

To show this progression in Figure \figthree{C}, we define $\Omega_{r}$
as the total weight on all edges of dimension $r$, and an effective
dimensionality $d_{\mathrm{eff}}=\sum_{r=1}^{D}r\,\Omega_{r}$. This
increases smoothly from 0 towards $D=2$ as the data improves.

\begin{figure}
\centering\includegraphics[width=0.95\columnwidth]{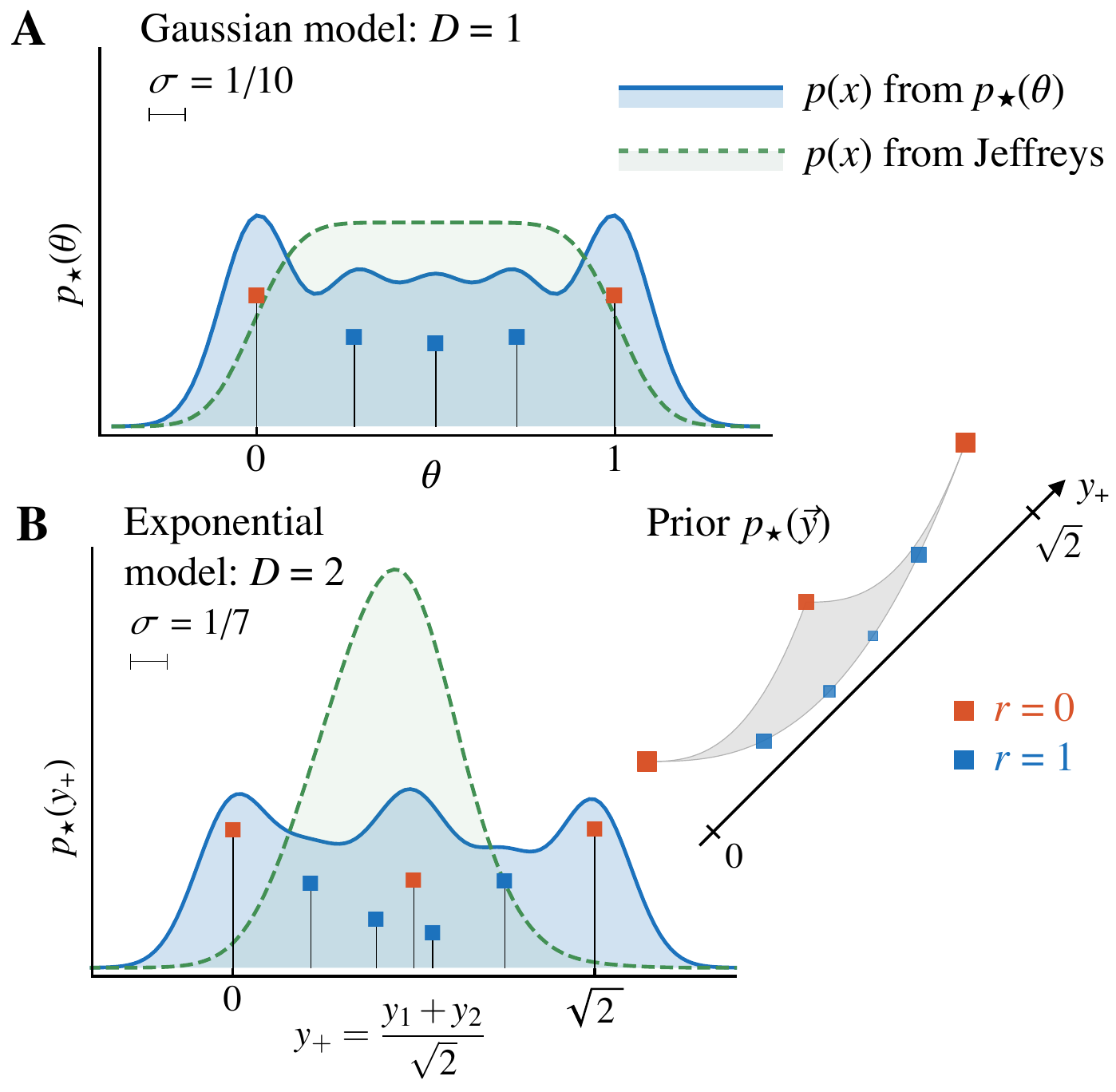}

\caption{\textbf{Distributions of expected data $p(x)$ from different priors.}
Panel A is the one-parameter Gaussian model, with $L=10$. Panel B
projects the two-parameter exponential model onto the $y_{1}+y_{2}$
direction, for $\sigma=1/7$ where the perpendicular direction should
be irrelevant. The length of the relevant direction is about the same
as the one-parameter case: $L_{+}=7\sqrt{2}$. Notice that the distribution
of expected data $p(x_{+})$ from Jeffreys prior here is quite different,
with almost no weight at the ends of the range ($0$ and $\sqrt{2}$),
because this prior still weights the area not the length. \label{fig:four} }
\end{figure}

At medium values of $\sigma$, the prior $\popt(\vec{y})$ almost
ignores the width of the parameter manifold, and cares mostly about
its length ($L_{+}=\sqrt{2}/\sigma$ along the diagonal). This behavior
 is very different to that of Jeffreys prior: in Figure \figfour{B}
we demonstrate this by plotting the distributions of data implied
by these two priors. Jeffreys puts almost no weight near the ends
of the long (i.e. stiff, or relevant) parameter's range, because
the (sloppy, or irrelevant) width happens to be even narrower there
than in the middle. By contrast our effective model puts significant
weight on each end, much like the one-parameter model in Figure \figfour{A}.

The difference between one and two parameters being relevant (in Figure
\figthree{B}) is very roughly $\sigma=1/7$ to $\sigma=1/50$, a
factor 7 in Fisher length, thus a factor 50 in the number of repetitions
$m$ \textemdash{} perhaps the difference between a week's data and
a year's. These numbers are artificially small to demonstrate the
appearance of models away from the boundary: more realistic models
often have manifold lengths spread over many orders of magnitude \cite{Gutenkunst:2007gl,Machta:2013ga},
and thus have some parameters inaccessible even with centuries of
data. To measure these we need  a qualitatively different experiment,
justifying a different effective theory. 

The one-dimensional model along the lower edge of Figure \figthree{A}
is the effective theory with equal decay constants. This remains true
if we allow more parameters $k_{3},k_{4},\ldots$ in \eqref{eq:forward-exp-model},
and $\popt(\vec{y})$ will still place a similar weight there.% 
\footnote{If we have more parameters than measurements then the model must be
singular. In fact the exponential model of Figure \ref{fig:three}
is already slightly singular, since $k_{1}\leftrightarrow k_{2}$
does not change the data; we could cure this by restricting to $k_{2}\geq k_{1}$,
or by working with $\vec{y}$, to obtain a regular model. } Measuring $x_{t}$ also at later times $t_{3},t_{4},\ldots$ will
add more thin directions to the manifold \cite{Transtrum:2011de},
but the one-dimensional boundary corresponding to equal decay constants
will still have significant weight. The fact that such edges give
human-readable simpler models (unlike arbitrary submanifolds) was
the original motivation for preferring them in \cite{Transtrum:2014hr},
and it is very interesting that our optimization procedure has the
same preference.\footnote{Edges of the parameter manifold give simpler models not only in the
sense of having fewer parameters, but also in an algorithmic sense.
For example, the Michaelis\textendash Menten model is analytically
solvable \cite{Schnell:1997gn} in a limit which corresponds to a
manifold boundary \cite{Transtrum:2014rma}. Stable linear dynamical
systems of order $n$ are model boundaries order $n+1$ systems \cite{Pare:2015bz}.
Taking some parameter combinations to the extreme can lock spins into
Kadanoff blocks \cite{Transtrum:2014rma}. }

\section*{Discussion}

 While the three examples we have studied here are very simple,
they demonstrate a principled way of selecting optimal effective theories, 
especially in high-dimensional settings. 
Following \cite{Sims:2006gu}, we may call this rational ignorance.

The prior $\popt(\theta)$ which encodes this selection is the maximally
uninformative prior, in the sense of leaving maximum headroom for
learning from data. But its construction depends on the likelihood
function  $p(x\vert\theta)$, and thus it contains knowledge about
the experiment through which we are probing nature. Jeffreys prior
$\pjeff(\theta)$ also depends on the experiment, but more weakly:
it is independent of the number of repetitions $m$, precisely because
it is the limit $m\to\infty$ of the optimal prior \cite{Clarke:1994gw,Scholl:1998kq}.

Under either of these prescriptions, performing a second experiment
may necessitate a change in the prior, leading to a change in the
posterior not described by Bayes' theorem. If the second experiment
is different from the first, then changing to Jeffreys prior for the
combined experiment (and then applying Bayes' rule just once) will
have this effect \cite{Lewis:2017uq,Lewis:2013bu}.\footnote{This view 
is natural in the objective Bayesian tradition, but see
\cite{Poole:2000ee} and \cite{Seidenfeld:1979cr,Kass:1996jj,Williamson:2009gq}
for alternative viewpoints.} Our prescription differs from Jeffreys in also regarding more repetitions
of an identical experiment as being different. Many experiments would
have much higher resolution if they could be repeated for all eternity.
The fact that they cannot is an important limit on the accuracy of
our knowledge, and our proposal treats this limitation on the same
footing as  the rest of the specification of the experiment. 

Keeping $m$ finite is where we differ from earlier work  on prior
selection. Bernardo's reference prior \cite{Bernardo:1979uq} maximizes
the same mutual information, but always in the $m\to\infty$ limit
where it gives a smooth analytically tractable function. Using $I(X;\Theta)$
to quantify what can be learned from an experiment goes back to Lindley
\cite{Lindley:1956bj}. That finite information implies a discrete
distribution was known  at least since  \cite{Farber:1967us,Smith:1971kt}.
What has been overlooked is that this discreteness is useful for avoiding
a problem with  Jeffreys prior on the  hyper-ribbon parameter spaces
generic in science \cite{Gutenkunst:2007gl}: because it weights the
irrelevant parameter volume, Jeffreys prior has strong dependence
on microscopic effects invisible to experiment. The limit $m\to\infty$
has erased the divide between relevant and irrelevant parameters,
by throwing away the natural length scale on the parameter manifold.
By contrast $\popt(\theta)$ retains discreteness at roughly this
scale,  allowing it to ignore irrelevant directions.  Along a relevant
parameter direction this discreteness is no worse than rounding $\theta$
to as many digits as we can hope to measure, and we showed that in
fact the spacing of atoms decreases faster than our accuracy improves. 

Model selection is more often studied not as part of prior selection,
but at the stage of fitting the parameters to data. From noisy data,
one is tempted to fit a model which is more complicated than reality;
avoiding such overfitting improves predictions. The AIC, BIC \cite{Akaike:1974ih,Schwarz:1978uv}
and related criteria \cite{Rissanen:1978ez,Wallace:1968co,Spiegelhalter:2002ii,Watanabe:2010uh,Watanabe:2013wy}
are subleading terms of various  measures in the $m\to\infty$ limit,
in which all (nonsingular) parameters of the true model can be accurately
measured. Techniques like MDL, NML, and cross-validation \cite{Rissanen:1978ez,Grunwald:2009ub,Arlot_2010}
need not take this limit, but all are applied after seeing the data.
They favor minimally flexible models close to the data seen, while
our procedure favors one answer which can distinguish as many different
outcomes as possible.  It is curious that both approaches can point
towards simplicity. We explore this contrast in more detail in the
Appendix.\footnote{Model selection usually starts from a list of models to be compared,
in our language a list of submanifolds of $\Theta$. We can also consider
maximising mutual information in this setting, rather than with an
unconstrained function $p(\theta)$, and unsurprisingly we observe
a similar preference for highly flexible simpler models. This is also
discussed in the Appendix, at \eqref{eq:MI-submanifold-comparison}. 
}

Being discrete, the prior $\popt(\theta)$ is very likely to exclude
the true value of the parameter, if such a $\theta_{\mathrm{true}}\in\Theta$
exists. This is not a flaw: the spirit of effective theory is to focus
on what is relevant for describing the data, deliberately ignoring
microscopic effects which we know to exist \cite{Anderson:1972dn}.
Thus the same effective theory can emerge from different microscopic
physics [as in the universality of critical points describing phase
transitions \cite{Batterman:2017bp}]. The relevant degrees of freedom
are often quasiparticles [such as the Cooper pairs of superconductivity
\cite{Bardeen:1957mv}] which do not exist in the microscopic theory,
but give a natural and simple description at the scale being observed.
We argued here for such simplicity not on the grounds of the difficulty
of simulating $10^{23}$ electrons, nor of human limitations, but
based on the natural measure of information learned.

There is similar simplicity to be found outside of physics. For example
 the Michaelis\textendash Menten law for enzyme kinetics \cite{%Menten:1913wn,
 Michaelis:2013bv}
 is derived as a limit in which only the ratios of some reaction rates
matter, and is useful regardless of the underlying system. In more
complicated systems which we cannot solve by hand, and for which the
symmetries and scaling arguments used in physics cannot be applied,
we hope that our information approach may be useful for identifying
the appropriately detailed  theory.

%\matmethods{ %%%%%%%%%%%%%%%%%%%%%%%%%%%%%%%%%%%%%%%
%
%methods??
%
%
%}    %%%%%%%%%%%%%%%%%%%%%%%%%%%%%%%%%%%%%%% end of methods
%
%\showmatmethods % Display the Materials and Methods section

%Please include your acknowledgments here, set in a single paragraph. 
% Please do not include any acknowledgments in the Supporting Information, or anywhere else in the manuscript.
\acknow{
We thank Vijay Balasubramanian, William Bialek, Robert de
Mello Koch, Peter Gr\"{u}nwald, Jon Machta, James Sethna, Paul Wiggins,
and Ned Wingreen for discussion and comments. We thank ICTS Bangalore
for hospitality.

H.H.M. was supported by NIH grant R01GM107103. 
M.K.T. was support by NSF-EPCN 1710727. 
B.B.M. was supported by a Lewis-Sigler Fellowship and by NSF PHY 0957573. 
M.C.A. was supported by NCN grant 2012/06/A/ST2/00396. 
}

\showacknow % Display the acknowledgments section

% \pnasbreak splits and balances the columns before the references.
% If you see unexpected formatting errors, try commenting out this line
% as it can run into problems with floats and footnotes on the final page.
%\pnasbreak

% Bibliography
\bibliographystyle{my-JHEP-4} %% knows about arxiv numbers
\bibliography{for-MI-auto,for-MI-manual}

%% number figures 
\setcounter{figure}{0}
\renewcommand\thefigure{S\arabic{figure}}    

%% number equations 
\setcounter{equation}{0}
\renewcommand\theequation{S\arabic{equation}}    

%% reset footnote counter
\setcounter{footnote}{0}

\begin{figure}
\centering \includegraphics[width=0.99\columnwidth]{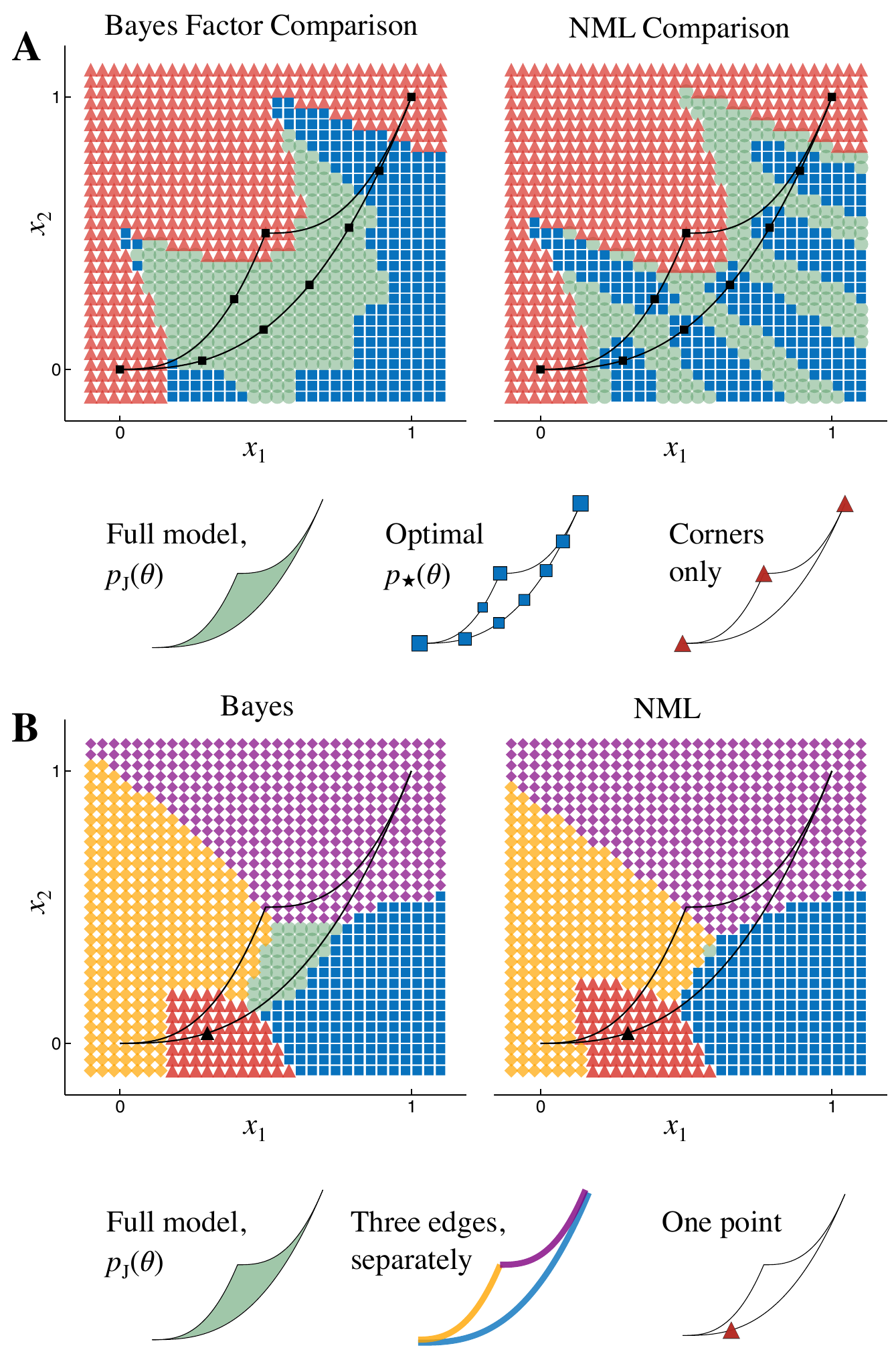}\caption{\textbf{Model selection from data point $x$.} Each large panel here
shows which of a list of effective models is preferred after observing
data $x\in X$. On the left the criterion is maximizing \eqref{eq:p-given-d},
on the right the criterion is \eqref{eq:p-NML}. We study the same
exponential model considered above, with $\sigma=1/10$. Panel A compares
our optimal effective model with prior $\popt(\theta)$ to the full
model (with Jeffeys prior) and to an even simpler model whose prior
is just three delta functions. (These are drawn in the legend below).
Panel B compares the full model to three different one-dimensional
models, each allowing only one edge of $\Theta$ (with a uniform prior
along this, i.e. the one-dimensional Jeffreys prior) and also to a
trivial model (just one point), again with colors as indicated just
below. \label{fig:six}}
\end{figure}

\section*{Appendix: Model Selection from Data}

The usual discussion of model selection takes place after
observing data $x$. If we wish to compare some models\footnote{The word model unfortunately means several things in the literature.
We mean parameter space $\Theta_{d}$ always equipped with a likelihood
function $p(x\vert\theta)$, and usually with a prior $p_{d}(\theta)$.
When this is a subspace of some larger model $\Theta_{D}$ (whose
likelihood function agrees, but whose prior may be unrelated) then
we term the smaller one an effective model, or a reduced model, although
we do not always write the adjective. The optimal prior $\popt(\theta)$
defines an effective model in this sense. Its support will typically
be on several boundaries of $\Theta_{D}$. If the boundaries of $\Theta_{D}$
(of all dimensions) are regarded as a canonical list of reduced models,
then $\popt(\theta)$ is seldom a sub-model of any one of them. } labeled by $d$, each with some prior $p_{d}(\theta)$, then one
prescription is to choose the model with the largest $p(x)$. Labelling
this explicitly, we write
\begin{equation}
p(x\vert d)=\int_{\Theta_{d}\negthickspace}d\theta\:p(x\vert\theta)\,p_{d}(\theta),\qquad p_{d}(\theta)>0\text{ on }\Theta_{d}\subset\Theta.\label{eq:p-given-d}
\end{equation}
If the Bayes factor $p(x\vert d)/p(x\vert d')$ is larger than one
then (absent any other information) $d$ is preferred over $d'$ \cite{Kass:1995iy}.\footnote{If one of the priors is improper, say $\int d\theta\:p_{d}(\theta)=\infty$,
then $p(x\vert d)$ will also be infinite. In this sense the Bayes
factor behaves worse than the posterior $p(\theta\vert x)$, which
can still be finite. } In the usual  asymptotic limit $m\to\infty$, this idea leads to
minimising the Bayesian information criterion (BIC) \cite{Schwarz:1978uv}:
\[
-\log p(x\vert d)\approx-\log p(x\vert\hat{\theta}_{d})+\frac{d}{2}\log m+\mathcal{O}(m^{0})
\]
where $-\log p(x\vert\hat{\theta}_{d})=\tfrac{1}{2}\chi^{2}=\tfrac{1}{2}\sum_{i=1}^{m}[x_{i}-y_{i}(\hat{\theta}_{d})]^{2}/\sigma^{2}\sim\mathcal{O}(m)$,
and $\hat{\theta}_{d}$ is a maximum likelihood estimator for $x$,
constrained to the appropriate subspace:
\[
\hat{\theta}_{d}(x)=\mathop{\mathrm{argmax}}_{\theta\in\Theta_{d}}p(x\vert\theta).
\]
The term $d\log m$ penalises models with more parameters, even though
they  can usually fit the data more closely. Despite the name this
procedure is not very Bayesian:  one chooses the effective model (and
hence the prior) after seeing the data, rather than simply updating
according to Bayes theorem.\footnote{Terms penalising more complex models can be translated into shrinkage
priors, which concentrate weight near to simpler models \cite{Bhadra:2016bc}.
Perhaps the shrinkage priors closest to this paper's are the penalised
complexity priors of \cite{Simpson:2017kw}. Those are also reparameterization
invariant, and also concentrate weight on a subspace of $\Theta$,
often a boundary. However both the subspace (or base model) and the
degree of concentration (scaling parameter) are chosen by hand, rather
than being deduced from $p(x\vert\theta)$.} 

\begin{figure}
\centering \includegraphics[width=0.95\columnwidth]{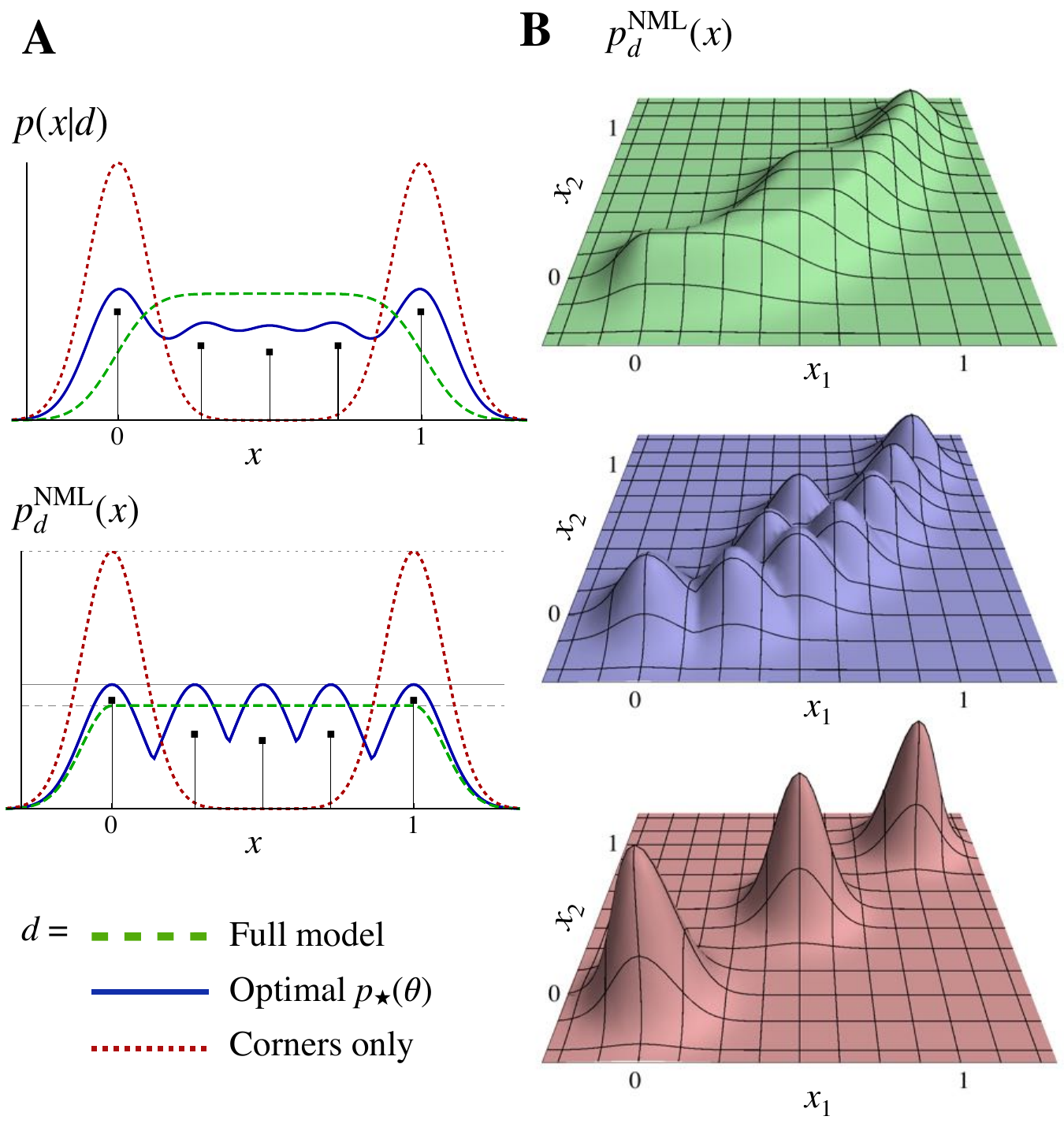}\caption{\textbf{Distributions over $X$.} Panel A shows $p(x\vert d)$ \eqref{eq:p-given-d}
and $p_{d}^{\mathrm{NML}}(x)$ \eqref{eq:p-NML} for the one-parameter
Gaussian example \eqref{eq:p-cond-gaussian} 
with $\sigma=1/10$.
The three models being compared are the full model (with a flat prior),
the simpler model defined by $\popt(\theta)$, and a model with just
the endpoints of the line. Under the Bayes factor comparison, the
$\popt(\theta)$ model would never be preferred here. Panel B draws
$p_{d}^{\mathrm{NML}}(x)$ for the two-parameter exponential model,
\eqref{eq:forward-exp-model}%
, for the complete model, the effective
model defined by the support of $\popt(\theta)$, and an even simpler
model allowing only three points --- the same three models as compared
in Figure \figsix{A}. Notice that $p_{d}^{\mathrm{NML}}(x)$ is always
a constant on the allowed region $x\in y(\Theta_{d})$.
\label{fig:nov} 
}
\end{figure}

Related prescriptions can be derived from minimum description length
(MDL) ideas. To allow reconstruction of the data we transmit both
the fitted parameters and the residual errors, and minimising the
(compressed) length of this transmission drives a tradeoff between
error and complexity \cite{Wallace:1968co,Rissanen:1978ez,Grunwald:2009ub}.
A convenient version of this goes by the name of normalized maximum
likeihood (NML) \cite{Myung:2006jl,Grunwald:2007vg}, and chooses
the model $d$ which maximizes 
\begin{equation}
p_{d}^{\mathrm{NML}}(x)=\frac{p\big(x\vert\hat{\theta}_{d}(x)\big)}{Z_{d}},\qquad Z_{d}=\int_{\Theta_{d}\negthickspace}dx'\:p\big(x'\vert\hat{\theta}_{d}(x')\big).\label{eq:p-NML}
\end{equation}
This is not Bayesian in origin, and does not depend on the prior on
each effective model $d$, only its support $\Theta_{d}$. The function
$p_{d}^{\mathrm{NML}}(x)$ is not expected data in the sense of $p(x)$
\textemdash{} it is not the convolution of the likelihood with any
prior.\footnote{This relevant optimization problem here can be described as minimizing
worst-case expected regret, written \cite{Myung:2006jl}
\[
p_{d}^{\mathrm{NML}}=\mathop{\mathrm{argmin}}_{q}\:\max_{x}\:\log\frac{p(x\vert\hat{\theta}_{d}(x))}{q(x)},\qquad\hat{\theta}_{d}(x)\in\Theta_{d}.
\]
Perhaps the closest formulation of our maximum mutual information
problem is that our $\popt(x)$, the distribution on $X$ not the
prior, can be found as follows \cite{Haussler:1997fa}:
\[
\popt=\mathop{\mathrm{argmin}}_{q\in B}\:\max_{\theta} {\textstyle\int_{X}}dx\:p(x\vert\theta)\log\frac{p(x\vert\theta)}{q(x)}
\]
where $q(x)$ is constrained to be a Bayes strategy i.e. to arise
from some prior $\popt(\theta)$. Note the absence of $\hat{\theta}_{d}(x)$
and the presence of an integral over $X$, corresponding to the fact
thats that this maximization takes place without being given a subspace
$\Theta_{d}$ nor seeing data $x$. The resulting distributions on
$X$ are also different, as drawn in Figure \ref{fig:nov}. If plotted
on Figure \figfour{B}, $p_{2}^{\mathrm{NML}}(x)$ from the full model
would be somewhere between the two expected data $p(x)$ lines there.
But it is not a comparable object; its purpose is model comparison
as in Figure \ref{fig:six}.} In the asymptotic limit $p_{d}^{\mathrm{NML}}(x)$ approaches $p(x)$
from Jeffreys prior, and this criterion agrees with BIC \cite{Rissanen:1978ez},
but away from this limit they differ. 

In Figure \ref{fig:six} we apply these two prescriptions to the exponential
example treated in the text. At each $\vec{x}\in X$ we indicate which
of a list of models is preferred.\footnote{Recall that $\vec{x}$ is $\vec{y}$ corrupted by Gaussian noise,
and $\vec{y}$ is constrained to the area show in Figure \figthree{A}
because it arises from decay rates $k_{\mu}$ via \eqref{eq:forward-exp-model}.
We may regard either $y_{t}$ or $k_{\mu}$ as being the parameters,
generically $\theta$. } Figure \ref{fig:nov} instead draws the distributions being used.
\begin{itemize}
\item Figure \figsix{A} compares three models: the complete model (with
Jeffreys prior), the optimal model described by discrete prior $\popt(\vec{y})$,
and an even simpler model with weight only on the three vertices $\vec{y}=(0,0)$,
$(\tfrac{1}{2},\tfrac{1}{2})$, $(1,1)$. 
\item Figure \figsix{B} instead compares the complete model to three different
one-parameter models (along the three boundaries of the allowed region
of the $\vec{y}$ plane) and a zero-parameter model (one point $\vec{y}$,
in no particularly special place). In terms of decay rates the three
lines are limits $k_{1}=k_{2}$, $k_{1}=0$ and $k_{2}=\infty$.
\end{itemize}
Different effective models are preferred for different values of data
$x$. At a given point $x$, if several models contain the same $\hat{\theta}(x)$
then the simplest among them is preferred, which in the NML case means
precisely the one with the smallest denominator $Z_{d}$. In fact
a trivial model consisting of just one point $\Theta_{0}=\hat{\theta}(x)$
would always be preferred if it were among those considered \textemdash{}
there is no automatic preference for models which can produce a wide
range of possible data. 

By contrast our prior selection approach aims to be able to distinguish
as many possible outcomes in $X$ as possible. Applied to the same
list of models as in Figure \ref{fig:six}, this gives the following
fixed scores (base $e$):
\[
I_{\text{full}}=1.296,\qquad I_{\star}=1.630,\qquad I_{\text{corners}}=1.098
\]
and 
\begin{equation}
\begin{gathered}I_{\text{upper}}=0.852,\qquad I_{\text{lower-left}}=0.845,\\
I_{\text{lower-right}}=1.418,\qquad I_{\text{one-point}}=0.
\end{gathered}
\label{eq:MI-submanifold-comparison}
\end{equation}
By definition $\popt(\theta)$ has the highest score. In second place
comes the line along the lower edge (corresponding to $k_{1}=k_{2}$).
The shorter lines are strongly disfavored, because they cover a much
smaller range of possible data. 

\begin{figure}
\centering \includegraphics[width=0.95\columnwidth]{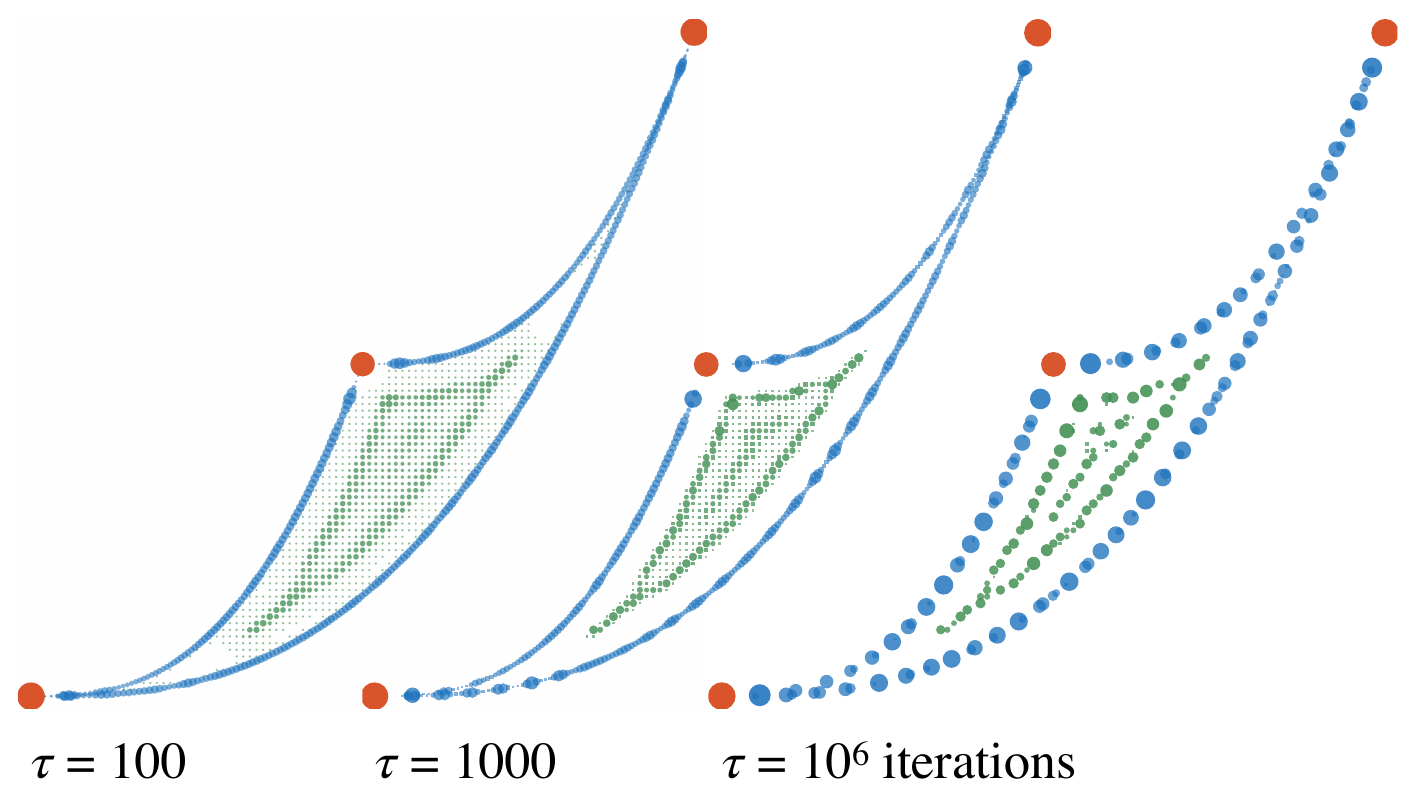} \caption{\textbf{Convergence of the BA algorithm (\ref{eq:BA-alg}) for the
exponential model.} This shows $p_{\star}(\vec{y})$ for the case
$\sigma=1/50$, with $\vec{y}$ discretized on a grid of spacing $1/100$
in the bulk and $1/200$ along the boundaries of the allowed region.
\label{fig:BA-convergence-2D}}
\end{figure}

\section*{Algorithms\label{sec:Algorithms}}

The standard algorithm for maximizing channel capacity (of
discrete memoryless channels) was written down independently by Blahut
\cite{Blahut:1972ed} and Arimoto \cite{Arimoto:1972jz}. This aspect
of rate-distortion theory is mathematically the same as the problem
we consider, of maximizing mutual information by choosing the prior.
The algorithm starts with $p_{0}(\theta)=\text{const.}$, and then
at each time step updates this by 
\begin{equation}
p_{\algtime+1}(\theta)=\frac{1}{Z_{\algtime}}e^{\fkl(\theta)}p_{\algtime}(\theta)\label{eq:BA-alg}
\end{equation}
where $Z_{\algtime}=\int d\theta'\:e^{\fkl(\theta')}p_{\algtime}(\theta')$
maintains normalization, and $\fkl(\theta)=D_{\mathrm{KL}}\left[p(x\vert\theta)\middle\|p(x)\right]$
is computed with $p_{\algtime}(\theta)$. Since this is a convex optimization
problem, the algorithm is guaranteed to converge to the global maximum.
This makes it a good tool to see discreteness emerging.

Figures \ref{fig:BA-new} 
and \ref{fig:BA-convergence-2D} show the
progress of this algorithm for the one- and two-dimensional models
in the text. We stress that the number and positions of the peaks
which form are unchanged when the discretization of $\theta$ is made
much finer. Notice also that the convergence to delta functions happens
much sooner near to the boundaries than in the interior. The convergence
to the correct mutual information $I(X;\Theta)$, and towards the
optimum distribution on data space $p(x)$, happens much faster than
the convergence to the correct number of delta functions.

Because $\theta$ must be discretized for this procedure, it is poorly
suited to high-dimensional parameter spaces. However once we know
that $\popt(\theta)$ is discrete it is natural to consider algorithms
exploiting this. With $\atoms$ atoms, we can adjust their positions
$\smash{\vec{\theta}_{a}}$ and weights $\lambda_{a}$ using gradients
\begin{align}
\frac{\partial\mathrm{MI}}{\partial\theta_{a}^{\mu}} & =\lambda_{a}\int dx\:\frac{\partial p(x\vert\vec{\theta})}{\partial\theta^{\mu}}\:\log\frac{p(x\vert\vec{\theta})}{p(x)}\:\bigg\rvert_{\vec{\theta}=\vec{\theta}_{a}}\nonumber \\
\frac{\partial\mathrm{MI}}{\partial\lambda_{a}} & =\fkl(\vec{\theta}_{a})-1.\label{eq:gradient-MI}
\end{align}
Figures \figone{}, \figtwo{A}, \ref{fig:four} 
and the square plot points in Figure \ref{fig:three} 
were generated this way. This
optimization is not a convex problem (there is some tendency to place
two atoms on top of each other, and thus use too few points of support)
but it can often find the optimum solution. We can confirm this by
calculating $\fkl(\theta)$ everywhere \textemdash{} any points for
which this is larger than its value at the atoms indicates that we
do not have the optimal solution, and should add an atom.

Monte Carlo algorithms for this problem have been investigated in
the literature, see \cite{Chang:1988bu,Lafferty:2001uj} and especially
\cite{Dauwels:2005vw}. (Incidentally, we observe that \cite{Chang:1988bu}'s
table 1 contains a version of scaling law \eqref{eq:scaling-law-rho},
with $\slope\approx1/2$. No attempt was made there to use the optimal
number of atoms, only to calculate the channel capacity to sufficient
accuracy.)

\end{document}